\documentclass[]{spie}  %>>> use for US letter paper
\usepackage{amsmath,amsfonts,amssymb}
\usepackage{graphicx}
\usepackage[colorlinks=true, allcolors=blue]{hyperref}

\title{Demonstrating the integration of a photonic lantern with an all-fiber-based nulling interferometer}

\author[a]{Jordan Diaz}
\author[a]{Rebecca Jensen-Clem}
\author[a]{Philip M. Hinz}
\author[a]{Daren Dillon}
\author[b]{Pradip Gatkine}
\author[a]{Aditya R. Sengupta}
\author[c]{Dan Sirbu}
\author[d] {Sarah Tedder}
\author[a]{Kevin Bundy}
\author[d] {Brian Vyhnalek}
\author[a]{Steph Sallum}
\author[a]{Matthew C. DeMartino}
\author[e]{Stephen Eikenberry}
\author[e]{Peter Delfyett}
\author[e]{Rodrigo Amezcua-Correa}
\affil[a]{Univ. of California, Santa Cruz, Santa Cruz, CA, 95064, USA}
\affil[b]{Univ. of California, Los Angeles, Los Angeles, CA, 90095, USA}
\affil[c]{NASA Ames Research Center, Moffett Field, CA, 94035, USA}
\affil[d]{NASA Glenn Research Center, Cleveland, OH, 44135, USA}
\affil[e]{Univ. of Central Florida, Orlando, FL, 32816, USA}

\authorinfo{Further author information: (Send correspondence to J.D.)\\ J. D.: E-mail: jdiaz71@ucsc.edu}

\begin{document} 
\maketitle

\begin{abstract}
High-contrast imaging of Solar System scale exoplanets and protoplanets demands advancements in instrumentation to access deeper starlight suppression at smaller angular separations than today's state-of-the-art. The multi-mode to single-mode conversion capabilities of photonic lanterns (PLs) provide new avenues to implement techniques such as nulling interferometry due to the inherent spatial filtering of single-mode waveguides. In this work, we present laboratory results on an all-fiber-based focal plane nulling interferometer using off-the-shelf components operating at 1550 nm. We demonstrate the implementation of a PL for coupling light into the instrument, and compare it to the case when laser light is directly fed into the interferometer. The integration of a PL with the interferometer evidences their potential for feeding photonic-based science instruments. Additionally, we discuss expanding the concept of the instrument for the detection of accreting protoplanets.
\end{abstract}

% Include a list of keywords after the abstract 

\section{INTRODUCTION}
\label{sec:intro}  % \label{} allows reference to this section
The number of directly imaged exoplanets has grown to a few dozen; however, these detections have been limited to massive, young, and hot planets at large separations from their host stars. Therefore, gaining insights into the formation and evolution of our solar system from these detections is difficult. In particular, expanding the population of directly detected protoplanets is crucial to advance our understanding of the processes of planet formation. At this date, only 3 protoplanets have been confirmed through the identification of H$\alpha$ flux emitted as they accrete material from the protoplanetary disk: PDS 70 b,c and Wispit 2b\cite{Keppler2018,Haffert2019,Close2025}. However, these systems are located at separations greater than or similar to those of the outermost planets of the Solar System. Hence, detecting Solar System analogs requires instruments capable of accessing higher contrasts at smaller angular separations than today's state-of-the-art.

Astrophotonics is a promising pathway towards achieving that goal. Photonic technologies offer potential benefits and advantages by enabling diverse functionalities within compact waveguide systems, such as flexible light guidance and rerouting, phase and amplitude modulation, and spectral, spatial and polarization filtering\cite{Jovanovic2023}. A technology of significant interest is the photonic lantern (PL): a tapered waveguide that transitions from a multi-mode (MM) waveguide at one end to an array of single-mode (SM) waveguides at the other end\cite{LeonSaval2010, Birks2015}. PLs have proven to be a powerful tool, finding various applications, including wavefront sensing\cite{Norris2020, Lin2025,Sengupta2026}, high-angular-resolution imaging\cite{Kim2025}, high-resolution spectroscopy\cite{Vievard2024}, and nulling\cite{Xin2022, Xin2024}. Moreover, PLs are key in enabling efficient coupling of MM telescope light and transforming it into SM light to feed photonic-based instruments.   

In Diaz \textit{et al.} 2024\cite{Diaz2024}, we introduced the concept of a focal plane all-fiber-based nulling interferometer comprised of off-the-shelf components. The instrument takes advantage of the inherent spatial filtering properties of SM fibers to improve the null and stabilizes it with active phase and intensity modulation components. By using a PL's ability to redistribute light to different ports depending on the object's position on-sky, a pair of ports containing light from an off-axis companion can be combined for suppressing the contaminating starlight. In the aforementioned work, we demonstrated a proof-of-concept in the laboratory of the fringe tracking and nulling capabilities of the instrument using visible laser light. Here, we present a second version of the instrument that operates in the infrared and demonstrate the integration of a PL to feed light into the instrument. Furthermore, we discuss the specific application of the instrument concept as a nulling interferometer dedicated to the identification of accreting protoplanets. 

\section{EXPERIMENTAL SETUP}
\label{sec:2}
\subsection{The nuller}
\label{{subsec:2-1}}
The IR version of the nulling interferometer, hereafter the ``nuller," builds on the design used in Diaz \textit{et al.} 2024\cite{Diaz2024} which follows that of a Mach-Zehnder interferometer. Here, one of the arms of the interferometer emulates the channel where only on-axis starlight is contained, and the other one where most of the off-axis planet light will be transmitted through with less contaminating starlight. In the former, an intensity modulation component is employed to reject most of the on-axis starlight and match it to the amount of contaminating starlight found in the latter, which is expected to be much less. The intensity of the light before the combination is monitored by sending 10\% of the light of each arm to reference photodetectors (PDs) and used to inform a Thorlabs variable optical attenuator (VOA) to adjust the intensity in the ``on-axis" arm. Light is combined with a 50:50 2x2 directional coupler (DC) resulting in a constructive interference output and a destructive interference output, the ``bright" and ``null" channels, respectively. The bright channel is used to do the fringe tracking and control an Evanescent Optics Inc. Model 915B fiber stretcher for fine phase modulation. A detailed description of the algorithms used to match the intensities in the arms and stabilize the null at a minimum can be found in Ref. ~\citenum{Diaz2024}.

The nuller sits on the bench where the SEAL\cite{JensenClem2021,JensenClem2025} and muirSEAL\cite{Sengupta2025} testbeds are also enclosed. The IR version of the nuller uses a Thorlabs KLS1550 Fabry-Perot diode laser with a central wavelength of 1550 nm with a maximum output power of $\sim$7 mW. The outputs are measured with Thorlabs DET08CFC PDs, and, as indicated in the following sections, a Thorlabs PM16-120 power meter was used. The output signals of the PDs are visualized and monitored as voltages with a Tektronix MDO3024 oscilloscope. Polarization adjustments are made using Thorlabs FPC560 three-paddle manual polarization controllers. All of the nuller's components employ SMF-28 type fiber. The components use FC/APC connectors to minimize back-reflections, with the only exception being the fiber patch cord coupled to the laser. More details will be provided in \S\ref{subsec:2-2}. 

The principles of operation of the VOA and the fiber stretcher were previously described in Ref.~\citenum{Diaz2024}. In a similar fashion, the response of the devices as a function of the voltage applied was characterized. For the VOA, a Thorlabs V1550A, a linear range of operation was found between 2-3 V, with a minimum change in the throughput of $\sim$0.506\% (within the linear range) set by the power supply's resolution of 1 mV. Similarly, for the fiber stretcher, the linear response $\Delta\phi=1.747\pi\Delta V$ was found, corresponding to physical length changes of $\Delta L=1.747\lambda\Delta V/(2n)$, where $n$ is the refractive index of the fiber's core. 

A major upgrade for the IR version is the addition of a Luna Innovations Inc. VLD-001 manual variable delay line (MVDL). The MVDL provides coarse phase modulation adjustments to minimize the path length difference between the arms of the nuller with a maximum delay of $\sim$20 cm and a readout scale resolution of 0.05 mm. 

%%%% FIGURE BLOCK DIAGRAM
\begin{figure} [h!]
   \begin{center}
   \includegraphics[width=15cm]{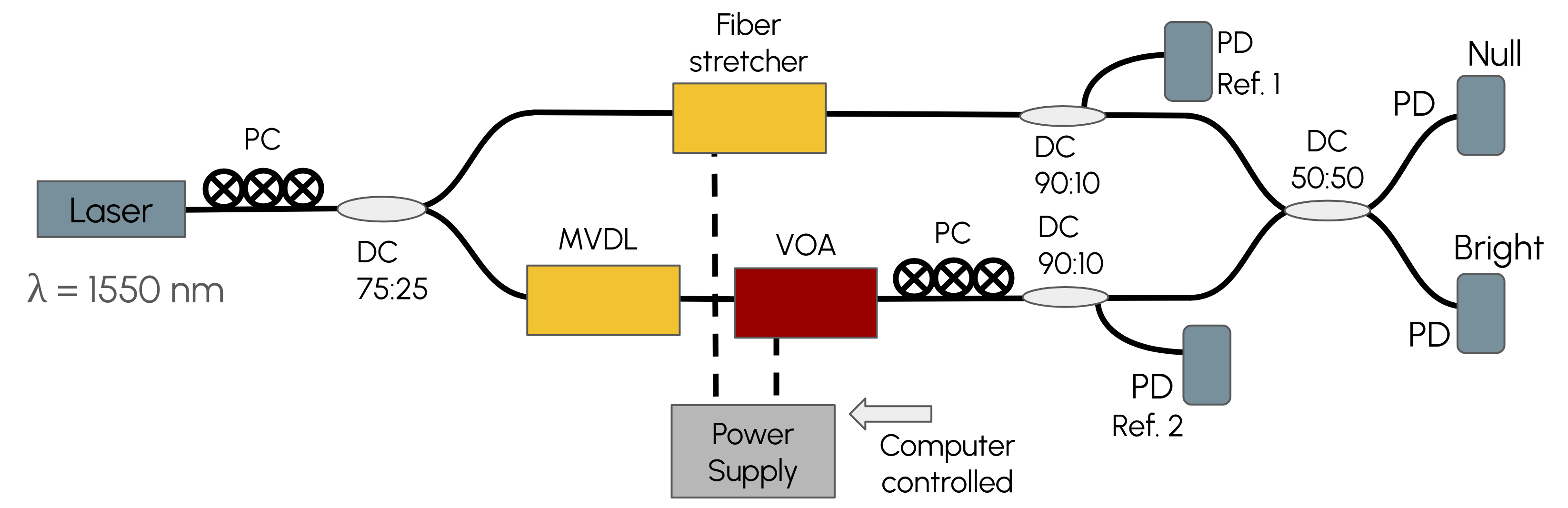}
   \end{center}
   \caption[example] 
   { \label{fig:setup_no_pl} 
The IR nuller configuration feeding light directly from the laser with a directional coupler (DC). MVDL-Manual variable delay line. VOA-Variable optical attenuator. PC-Polarization controller. PD-Photodetector.}
\end{figure} 
%%%% FIGURE BLOCK DIAGRAM
\begin{figure} [h!]
   \begin{center}
   \includegraphics[width=16cm]{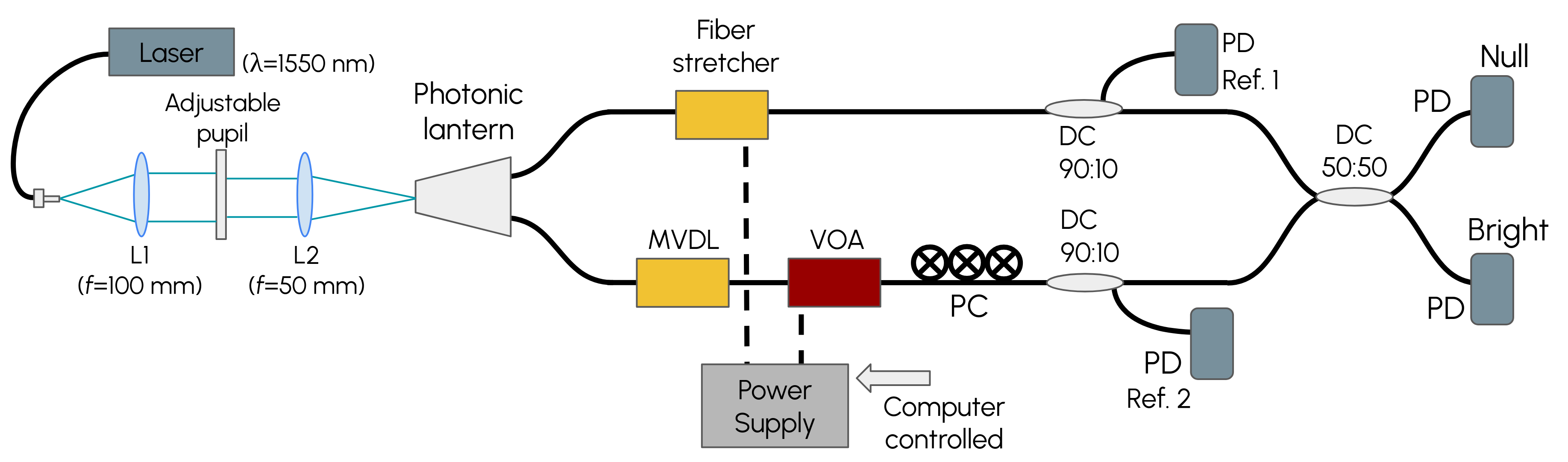}
   \end{center}
   \caption[example] 
   { \label{fig:setup_pl} 
The IR nuller configuration using a PL to feed light into the instrument.}
\end{figure} 
\subsection{Feeding light into the nuller}
\label{subsec:2-2}
In this work, two configurations to feed light into the nuller were investigated: 1) using a directional coupler directly connected to the laser and 2) using two output ports from a PL, shown respectively in Fig. \ref{fig:setup_no_pl} and Fig. \ref{fig:setup_pl}. In the first configuration, light from the laser is coupled into a hybrid fiber patch cord, with a FC/PC connector interfaced directly to the laser and a FC/APC connector at the other end connected to a PC. The PC then couples into a 75:25 DC, which feeds the nuller. 

The second configuration employs a PL to feed the nuller. For this, we used the injection unit to couple light into the PL shown in Fig. \ref{fig:setup_pl}. In contrast to the previous configuration, the fiber patch cord interfaced with the laser is not hybrid, and both of its connectors are FC/PC. The beam coming from the fiber is collimated and passes through an adjustable pupil to modify the $f/\#$ of the beam coupling into the PL's MM end. The PL is mounted on a 5-axis manual kinematic mount placed on an optical rail, enabling translation on the Z-axis to adjust the focus on the MM end of the PL. The SMF outputs of the PL were fixed to two quadruple L-bracket mating sleeves. From this, the input arms of the nuller were connected to the desired PL output ports to be used in our experiments.
\subsection{Characterization of the photonic lantern}
\label{subsec:2-3}
\begin{figure} [h!]
   \begin{center}
   \includegraphics[width=17cm]{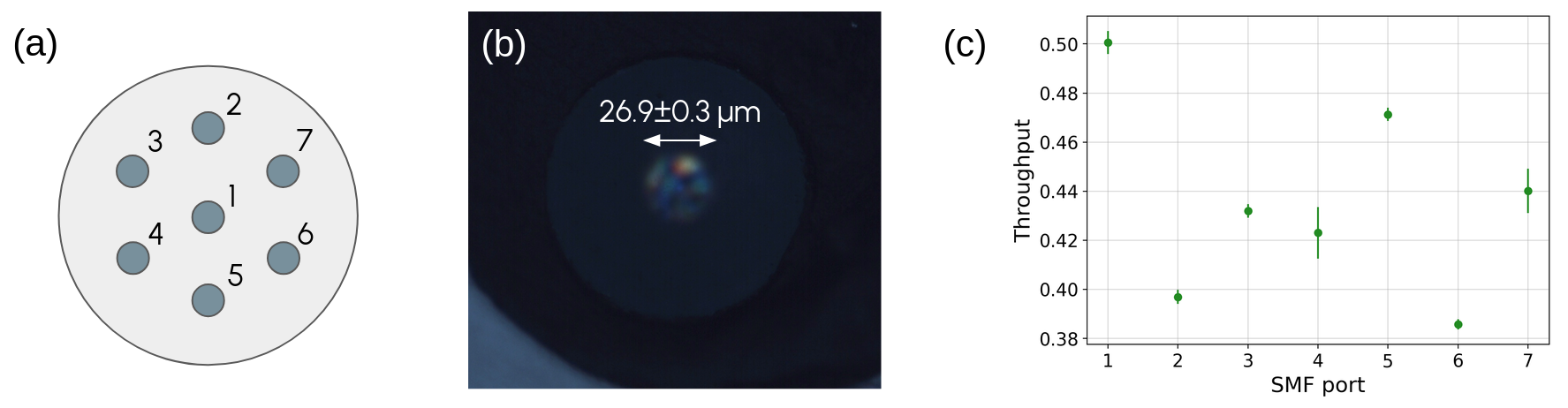}
   \end{center}
   \caption[example] 
   { \label{fig:PL_char} 
(a) Labeling of the PL's SMF outputs used in the experiments described in this work. (b) Microscope image of the PL's MM end, showing the MM intensity structure when light is coupled into a SMF. (c) End-to-end characterization of the throughput from the SMFs to the MM port measured by back illuminating the SMF ports with the laser.}
\end{figure}

The PL used in our experiments was borrowed from the UCLA AstroPhotonics Technology Lab (PI: P. Gatkine). The PL has 7 SMF pigtails differing in length and which were labeled as shown in Fig.\ref{fig:PL_char}(a). The MM end of the PL was imaged using an Olympus STM6-F10-3 measuring microscope. The SMFs were back illuminated individually with a white light flashlight, resulting in an intensity pattern at the MM core like the one depicted in Fig.\ref{fig:PL_char}(b). From the images, we measured a MM core diameter of 26.9$\pm$0.3 $\mu$m.

The end-to-end throughput of the SMF ports to the MM port was measured. Each SMF was directly fed with the laser, and the power at the output was measured with the power meter. The optical power going into the SMFs and at the output was measured three times. Figure \ref{fig:PL_char}(c) shows the mean value of the throughput for each port with their corresponding standard deviations, which were $<$2\%. The highest throughput of $\sim$50\% measured sets a limit for the amount of light that can be transmitted to that SMF port when light is coupled at the MM end, that limit being $\sim$7.15\% (assuming equal distribution across the 7 ports). 

%%%% FIGURE BLOCK DIAGRAM

\section{MEASUREMENTS OF THE NULL DEPTH}
\label{sec:3}
The primary goal of this work was to validate the implementation of  a PL as a tool to enable coupling of MM light to feed a photonic SM instrument downstream, particularly, in the context of nulling interferometry. For this reason, we conducted experiments to compare the light rejection performance of the nuller under the two light-feeding configurations described in \S~\ref{subsec:2-2}.   
 
\subsection{Case 1: feeding the nuller with the directional coupler}
\label{subsec:3-1}
\begin{figure} [h!]
   \begin{center}
   \includegraphics[width=15cm]{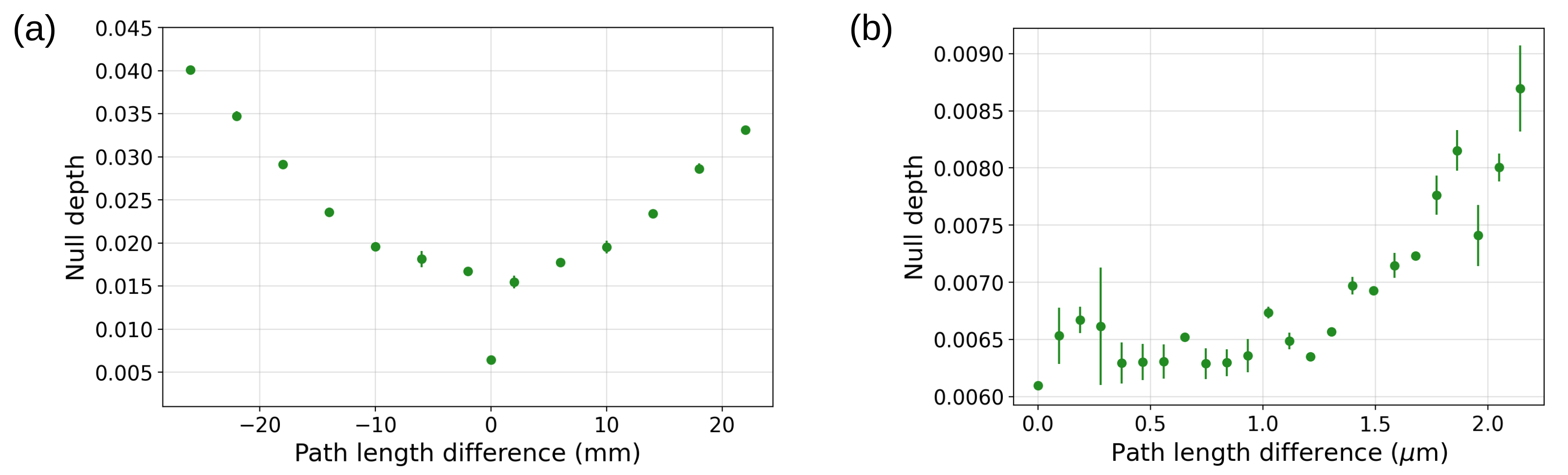}
   \end{center}
   \caption[example] 
   { \label{fig:Null_no_PL} 
Null depth measurements of the ``direct light injection" (no PL) configuration. The null depth is mapped using (a) coarse PLDs with the MVDL and (b) fine PLDs with the fiber stretcher.}
\end{figure}

The intensity of the light in the arms of the nuller was matched, and, using the PC downstream the laser and the second PC in the VOA arm, the polarization was adjusted to maximize the amplitude of the interference observed in the oscilloscope. Using the MVDL, the path length difference (PLD) between the arms was modified, and the minimum and maximum of the interference signal was recorded for each position. The polarization was constantly checked and corrected for variations induced by fluctuations in the environment. Figure \ref{fig:Null_no_PL} (a) shows the null depth as a function of PLDs introduced with the MVDL. Here, the position of PLD$\,=0$ corresponds to the PLD where the deepest null depth was observed for this set of measurements. The null depth traces a parabolic-shaped envelope, gradually decreasing as the PLD is reduced, however, it decreases abruptly nearly by a factor of 2 when the path lengths of the arms are closely matched. When the MVDL was set around this region, the interference exhibited a high sensitivity to external perturbations. 

Once the MVDL was positioned where the amplitude of the interference was maximized, a fine sweeping around this region was conducted using the fiber stretcher. The fiber was stretched by feeding it voltages in steps of 0.1 V, corresponding to physical path length differences of $\sim$0.135 $\mu$m. We recorded the power in one of the output channels with the power meter, taking two samples of 20 s each, ensuring that for each sample a minimum and a maximum of the interference was recorded. The results are displayed in Fig. \ref{fig:Null_no_PL}(b). We note that the null depth was not improved within the swept PLD range, and deteriorated as the PLD increased. This degradation, while it is expected as the PLD increases, can also be potentially attributed to temperature and mechanical drifts over the course of the experiment that shifted the zeroth order interference fringe a few wavelengths off from its original position during the first measurement. 

%%%% FIGURE BLOCK DIAGRAM
%%%%%%
\subsection{Case 2: feeding the nuller with the photonic lantern}
\label{subsec{3-2}}
A similar process was followed with the PL. Given the large number of possible combinations of SMF ports to interface with the nuller (21 combinations), we focused on exploring three different scenarios: 1) combining the central port (port 1) with a port from the outer ring (ports 2-7), 2) combining two ports from the outer ring, and 3) using a port combination from scenario 1) but changing the beam's $f/\#$. Since the PL SMFs' lengths varied between $\sim$7-23 cm, we narrowed the selection of ports to those with the smallest length differences, this would ensure that we could make the most use of the MVDL delay range. Ports 1, 2 and 4 were selected for our experiments. 

\begin{figure} [h!]
   \begin{center}
   \includegraphics[width=8cm]{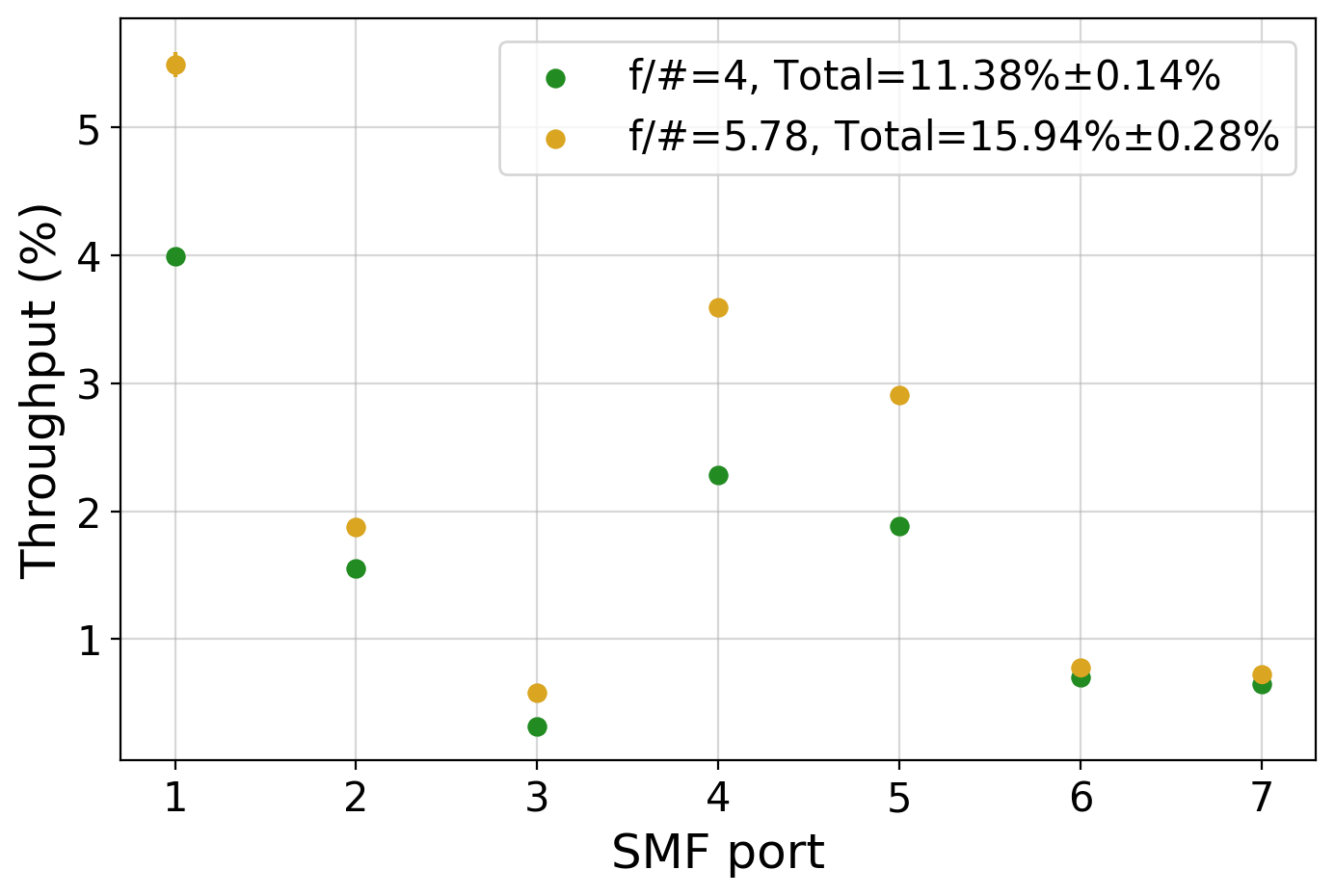}
   \end{center}
   \caption[example] 
   { \label{fig:throughput_fw} 
Throughput measurements of the PL's SMF outputs for the two $f/\#$s used in the null depth measurements.}
\end{figure}

Before connecting the PL ports to the nuller, the coupling of the laser light into the PL was optimized with the 5-axis stage and the translation stage on the optical rail while the output power at the central port was constantly monitored using the power meter. The $f/\#$ was originally set to 4. The pupil size was changed, finding the highest throughput values around $f/\#\sim4-6$, a behavior consistent with previous simulations and experimental measurements of throughput in lanterns\cite{Tedder2019, Lin2022a, Sengupta2025}. For our experiments, we used $f/\#=4$ and 5.78. For both $f/\#$s, we measured the power at the output of the SMFs and before the MM end. Figure \ref{fig:throughput_fw} shows the throughputs for each SMF port and the total throughputs. 

%%%% FIGURE BLOCK DIAGRAM
%%%%%%
After the adjustment of the coupling, the two ports selected for the scenario under investigation were interfaced with the nuller arms, and their intensities were matched using the VOA. The outputs were monitored with the oscilloscope, and the amplitude of the interference was maximized using the PC. A sample of the power coming from the nuller outputs was recorded with the power meter, again ensuring that for each sample a minimum and a maximum of the interference was recorded. For each measurement, a sample of 25-30 seconds was recorded. %with the Thorlabs \textit{Optical Parameter Monitor} software. 
The measurements were repeated over an extended range of PLDs controlled with the MVDL. The polarization was constantly checked and adjusted as required.

\begin{figure} [h!]
   \begin{center}
   \includegraphics[width=10cm]{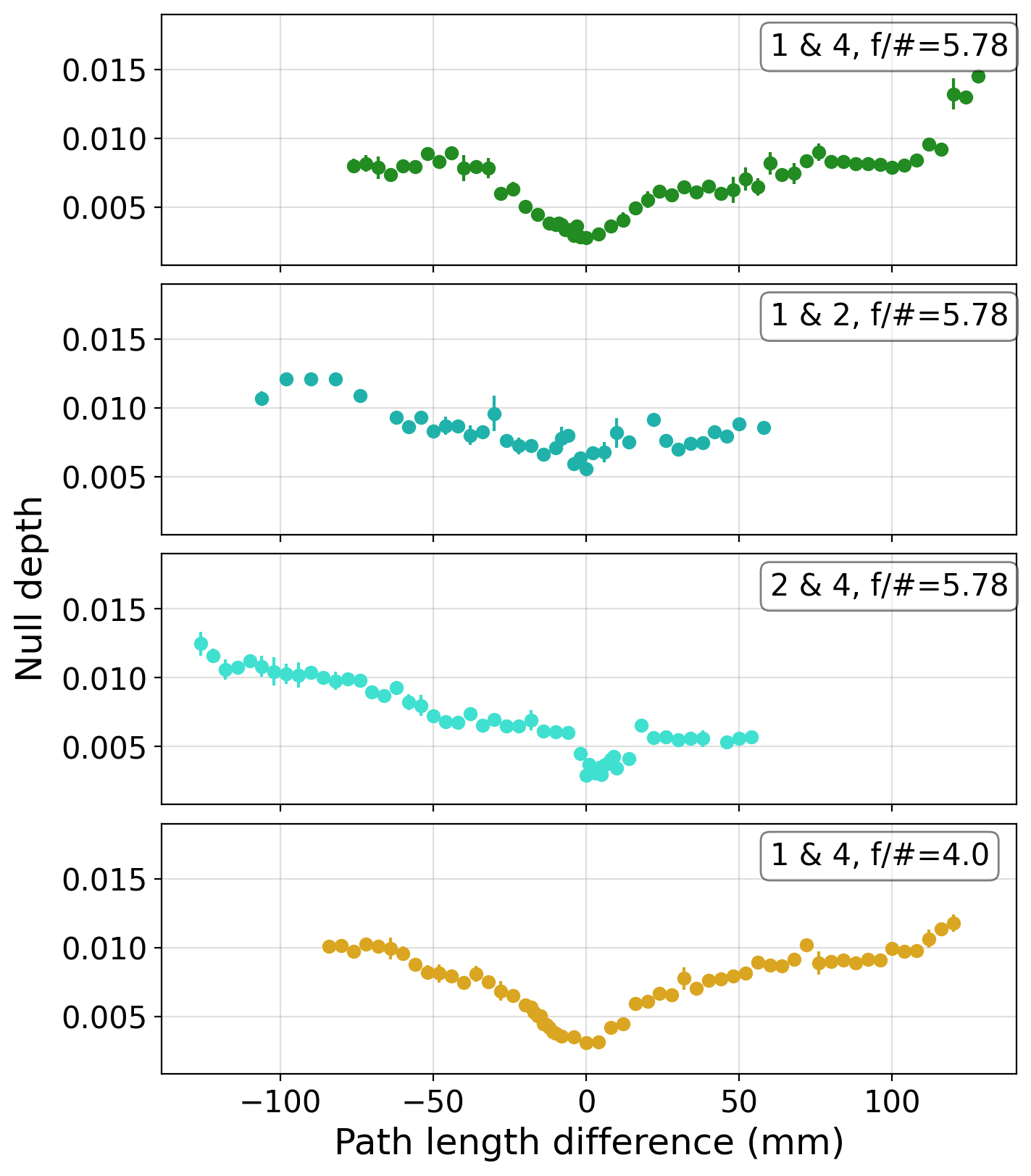}
   \end{center}
   \caption[example] 
   { \label{fig:Null_PL} 
Null depth measurements of the PL configuration for the three scenarios under investigation. The PLDs where introduced using the MVDL.}
\end{figure}

Figure \ref{fig:Null_PL} shows the results of the three scenarios explored. Two aspects can be highlighted from these results: 1) the envelope of the null depth is consistent across the three scenarios, with no substantial differences in their shape and depth, and 2) the shape of the envelope is flatter and exhibits a deeper null level over an extended range of PLDs compared to the non-PL case (see Fig. \ref{fig:Null_no_PL}(a)). The dip around the central region is observed again,but this time it drops off less dramatically. Also, the maximum level of light rejection reached was slightly improved by a factor of $\sim$2-3 ($N\approx2\times10^{-3}$).

%%%% FIGURE BLOCK DIAGRAM
%%%%%%
\subsection{Modeling the expected null depth}
\label{subsec:3-3}
Figures \ref{fig:Null_no_PL} and \ref{fig:Null_PL} reveal a notable discrepancy in the performance of the nuller between the two methods used to feed light into it. To explain this discrepancy, we investigate how the null depth in each case compares with theoretical expectations. First, we compute the expected null depth as a function of the PLD between the arms of the interferometer. The null depth can be calculated by integrating the residual intensity of the constructive interference for a given bandpass $BW$ as
\begin{equation}
   N\left(\lambda\right) = \int_{\lambda_{0}-BW/2}^{\lambda_{0}+BW/2} \frac{1 + \cos\left(\Phi(\lambda) \right)}{2BW} \,d\lambda, \label{eq:1}
\end{equation}
where $\lambda_{0}$ is the central wavelength\cite{Hinz2021}. The phase difference (in waves) between the two beams is given by 
\begin{equation}
    \Phi\left(\lambda\right) = \frac{n\lambda_{0}}{\lambda}+\frac{1}{2}, \label{eq:2}
\end{equation}
where $n$ is the number of waves off from the zero-order interference fringe.

Equation \ref{eq:1} assumes that all wavelengths within the bandpass of integration contribute equally to the residual intensity; however, the amount of light contributed by each will depend on the emission spectrum of the source. For the laser used in our experiments, no public documentation containing information about its spectral emission profile was found. The manufacturer provided a value of the linewidth $\Delta\lambda=70$ pm through private communication, but it was indicated that this was an estimate that may differ from the actual value. A way to assess the accuracy of this estimate of the linewidth consists in the analysis of the visibility of the interference fringes. The Wiener-Khintchine theorem relates the visibility of interference fringes to the temporal coherence of a source and, consequently, to its coherence length $L_{coh}$\cite{Goodman2015, Akcay2002}. For a source with a Lorentzian spectrum, its coherence length corresponds to the PLD between the arms of the interferometer where visibility drops to $1/e$. The coherence length is related to the linewidth of the source as
\begin{equation}
    L_{coh} = c\tau_{coh} = \frac{c}{\pi\Delta\nu} = \frac{\lambda_{0}^{2}}{\pi \Delta\lambda}, \label{eq:3}
\end{equation}
where $c$ is the speed of light and $\tau_{coh}$ is the coherence time\cite{Goodman2015}. 

\begin{figure} [h!]
   \begin{center}
   \includegraphics[width=8cm]{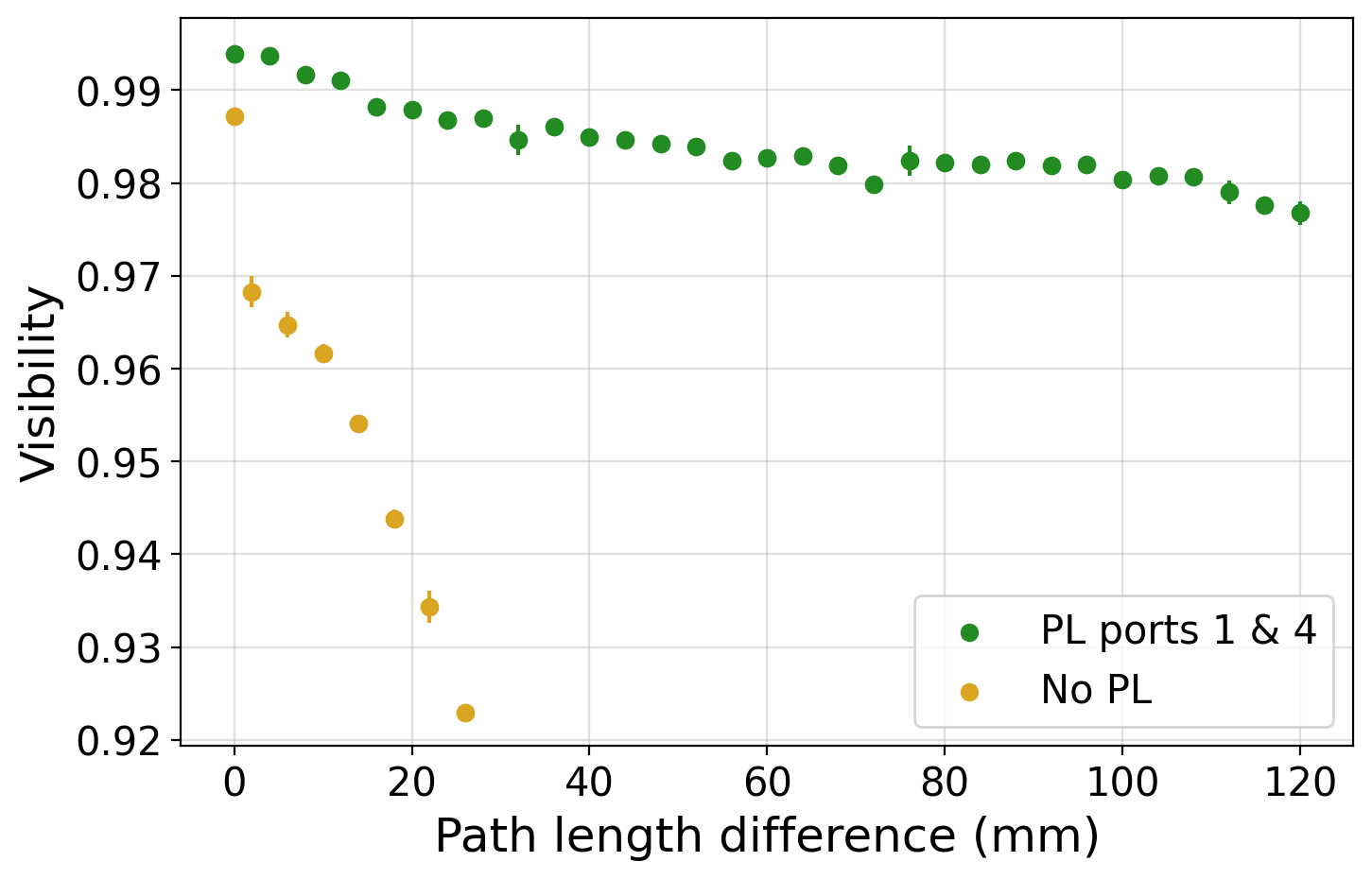}
   \end{center}
   \caption[example] 
   { \label{fig:Visibilities} 
Visibility of the interference fringes as a function of PLD for the two light-injection configurations investigated.}
\end{figure}

We calculated the visibility as a function of PLD when the nuller was fed with the directional coupler and when ports 1 and 4 of the PL were used as shown in Fig.\ref{fig:Visibilities}. We note that even for the non-PL case, the visibility is $>$90\% for PLDs$>$2 cm, differing from the expected coherence length of $\sim$1 cm when $\Delta\lambda=70$ pm. The visibilities measured suggest that the coherence length should be at least an order of magnitude larger, and, in turn, the linewidth an order of magnitude smaller. While this provides partial information on the light source's characteristics, details about its emission spectrum are missing. 

%%%% FIGURE BLOCK DIAGRAM
\begin{figure} [h!]
   \begin{center}
   \includegraphics[width=17cm]{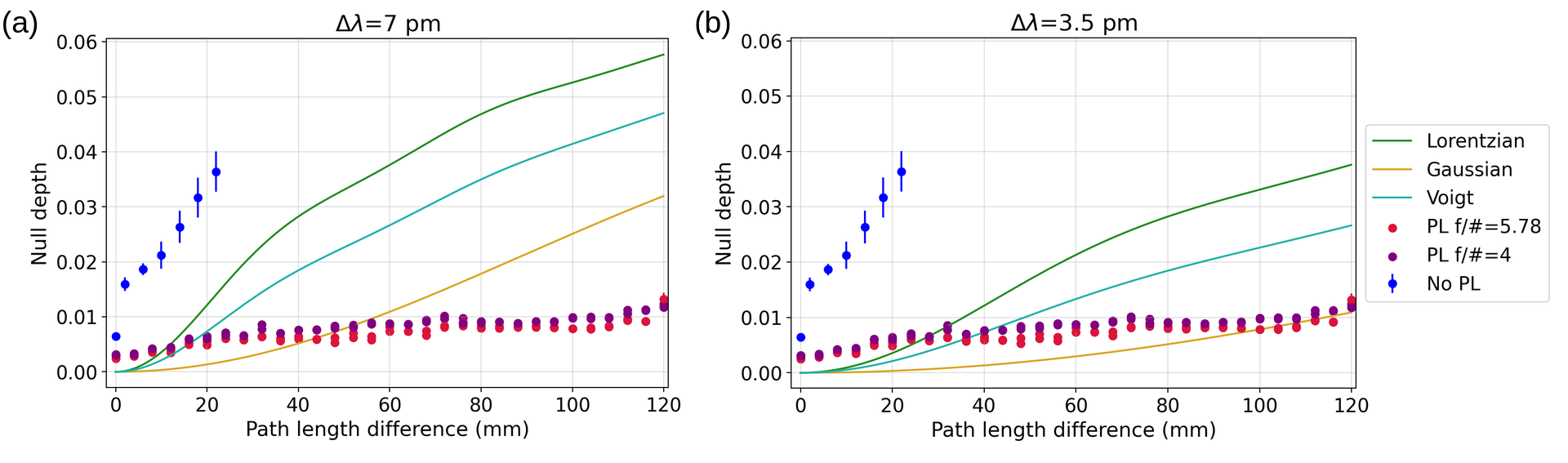}
   \end{center}
   \caption[example] 
   { \label{fig:null_depth_models} 
Curves of the expected null depth as a function of PLD when different emission profiles of the laser are assumed. In both cases, a bandpass of 10$\Delta\lambda$ is considered. The models are compared to experimental measurements of the two configurations.}
\end{figure}

%%%%%%
Using Eq. \ref{eq:1}, we computed the null depth as a function of PLD and scaled it by the three typical spectral lineshapes of lasers: Lorentzian, Gaussian, and Voigt profiles. Figure \ref{fig:null_depth_models} displays the models when the profiles have a linewidth of 7 pm and 3.5 pm and the integration is done over a bandpass of $10\Delta\lambda$. The experimentally measured null depths for the non-PL case and the PL case (ports 1 and 4, $f/\#=4$) are also portrayed for comparison. The preliminary models fail to reproduce the experimental data due to the limited knowledge of the laser's emission characteristics. Nevertheless, they provide some hints about the expected behavior of the null depth: if the laser's linewidth is at least an order of magnitude smaller as suggested by the visibility, $\Delta\lambda<7$ pm, the null depth resulting from the PL case is in better agreement with the anticipated behavior and levels of the null depth.

%%%% FIGURE BLOCK DIAGRAM
%%%%%%
\subsection{Potential causes of discrepancy}
We hypothesize that the observed discrepancy can be caused by an error in the light injection when the directional coupler is used instead of the PL, particularly at the interface between the laser and the fiber patch cable connected directly to it. A possible explanation for this is that a back reflection is generated and is propagated through the system. The back reflection would acquire a delay long enough to make it lose its coherence with respect to the non-reflected beam, thus no longer meeting the destructive interference criterion and leaking into the null depth. 

Different routes can be followed to address the inconsistency between these results. A first approach is characterizing the laser's spectrum and computing higher fidelity curves of the null depth to compare with our current measurements. This can be done by measuring the spectrum directly with an optical spectrum analyzer or by measuring the visibilities over a larger range of PLDs. Alternatively, a different laser could be used. While using a source with a characterized spectrum is preferred, as it would enable computing more accurate models of the expected null depth, we would still benefit from using a non-characterized source by  corroborating whether the discrepancy is still observed between the light-feeding methods. Also, including a circulator downstream of the laser can prevent any back reflections from leaking into the system. A fourth approach, highly relevant to the goal of this work, is using another PL. Integration of a PL with a different design (e.g. different number of ports) would corroborate the observed performance of the nuller with the PL and would be informative in understanding the practical limitations and impact of PL design for the nulling. 

\subsection{Limitations}
\label{subsec:3-4}
Although our results validate the integration of the PL with the nuller, there is still room to improve the performance of this system. Nulling with the PL outperformed the case when the directional coupler fed the nuller by maintaining deeper levels of light rejection over extended PLD ranges and reaching a maximum null depth $\sim$3 times deeper ($2\times10^{-3}$ with the PL and $6\times10^{-3}$ with the directional coupler). However, the current setup only has polarization control in one of the arms of the interferometer (see Figs. \ref{fig:setup_no_pl} and \ref{fig:setup_pl}), which prevents optimal polarization matching and degrading the null. Including a second polarization controller to match the polarization before beam combination will potentially enable deeper nulls. 

Although in this work we do not present results on the stabilization of the null through active control, this functionality is limited in speed by the power supply controlling the fiber stretcher. In Diaz \textit{et al.} 2024\cite{Diaz2024}, we demonstrated fringe tracking control limited to speeds $<$10 Hz with the same device. Higher speeds can be achieved by implementing a field-programmable gate array (FPGA) to run the control loop\cite{Birbacher2024}.

\section{FUTURE WORK}
\label{sec:4}
\begin{figure} [h!]
   \begin{center}
   \includegraphics[width=14cm]{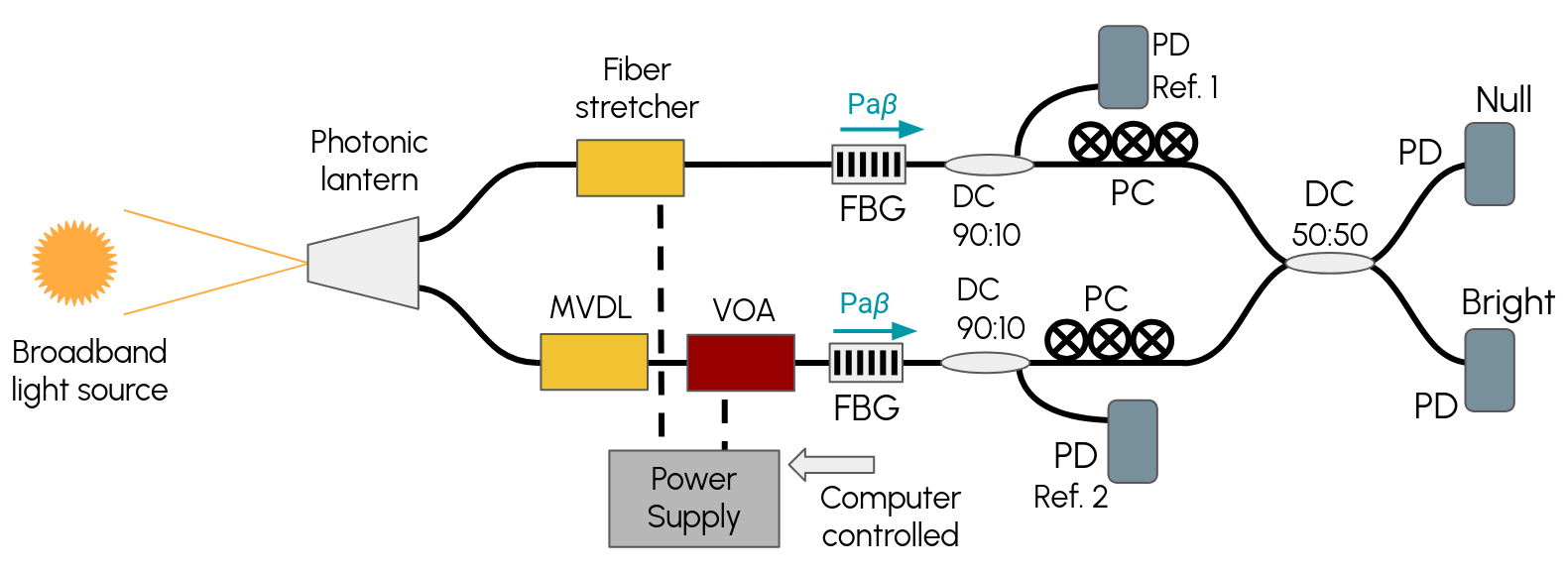}
   \end{center}
   \caption[example] 
   { \label{fig:Setup_FBG} 
Future configuration of the nuller to operate at the Pa$\beta$ accretion emission line wavelength. Two fiber bragg gratings (FBGs) are incorporated into the arms of the nuller to filter out other wavelengths.}
\end{figure}

Beyond performing the light rejection functionality of a nulling interferometer, our goal with the nuller is to develop an instrument devised for the detection of accreting protoplanets. In order to do that, the operating wavelength would be narrowed down to one of the hydrogen recombination lines known as tracers of accretion; here we consider Pa$\beta$\cite{Betti2022, Marleau2024}. Now that we have validated the integration of a PL to feed the nuller, a broadband light source will be used. Fiber bragg gratings (FBGs) will be incorporated downstream of the phase and intensity modulation components to transmit only the Pa$\beta$ light for nulling. Figure \ref{fig:Setup_FBG} shows the schematic layout that illustrates the future version of the setup. The rejected light at other wavelengths could be used for other purposes, such as wavefront sensing.

Before proceeding to move to the next stage of the nuller, we will first address the discrepancy presented earlier through one, or more, of the approaches proposed in \S\ref{subsec:3-3}. Our group will have access to a PL with 19 SMF pigtails in the near future, for which we will prioritize the alternate PL approach. Another aspect that will need to be studied as we move forward with the integration of PLs is the characterization of how light is distributed to the SMFs as a function of the beam's coupling position at the MM end. Coupling maps of PLs have been experimentally measured for high resolution imaging using the spectroastrometry technique\cite{Kim2024,Kim2025}. For the case of our nuller, these coupling maps will be necessary to select the ports that would confine most of the planet's light (as a function of a planet's separation and position angle) and interfere them. Another aspect to be considered is the chromaticity of the PL, which can be assessed by measuring coupling maps at different wavelengths with a monochromator.
%%%% FIGURE BLOCK DIAGRAM
%%%%%%
\section{SUMMARY}
\label{sec:5}
We have demonstrated the application of a 7-port photonic lantern to feed an all-fiber-based nulling interferometer using monochromatic light in the NIR ($\lambda=1550$ nm). The setup of the nuller presented here is an upgrade of the instrument introduced in Ref.~\citenum{Diaz2024}, with the capability of coarse phase modulation via a manual variable delay line. We compared two configurations for sending light into the nuller: 1) feeding light directly from a laser with a directional coupler and 2) by coupling light into a PL and using a pair of its SMF ports to feed it. We measured the null depth as a function of PLD for both configurations and found a notable discrepancy: the null was less sensitive to PLDs when the PL was used, exhibiting deeper null levels over an extended range of PLDs and improving the deepest null reached by a factor of 3. We computed preliminary models of the expected null depth limited by the partially known emission characteristics of the laser utilized, which suggest that the observed null depths for the PL case are closer to what should be expected. We hypothesize that this unexpected difference may be caused by a back reflection arising at the interface between the laser and the optical fiber system, thus losing its coherence and leaking into the null output. 

Future work will include addressing the discrepancy between the two configurations used by one or more of the approaches proposed in \S\ref{subsec:3-3}. Testing with a different PL will be explored in the near future, which will not only help provide more evidence to our findings but will also enrich our understanding of the benefits and limitations of combining these devices with SM instruments. Furthermore, the next step towards developing the nuller as an instrument for detecting accreting protoplanets will consist of feeding it with broadband light and conducting the nulling with Pa$\beta$ light. For this, the nuller will be upgraded by incorporating customized FBGs to separate out light with a wavelength of $\lambda=1282$ nm. 

Certainly, PLs have been applied for nulling, however, in these studies, the nulling is performed by the lantern itself by exploiting the spatial symmetry of mode-selective PLs to cancel out the on-axis starlight\cite{Xin2022, Xin2024}. The work presented here is, to the best of the authors' knowledge, the first to validate the application of a PL for nulling interferometry with active phase and intensity control components. Besides developing the case of an instrument for the identification of accreting companions, this and future work will contribute to the advancement of photonic-based systems working in synergy with PLs, such as photonic integrated circuits, for future instruments for ground- and space-based telescopes.

\acknowledgments % equivalent to \section*{ACKNOWLEDGMENTS}       
 
We gratefully acknowledge the University of California Observatories for funding this research. J.D. acknowledges the Cota-Robles Fellowship and the UCSC Community Connections group for their support. J.D. is also thankful to Ruslan Belikov and Kevin Fogarty for valuable discussions on this project.

% References
\bibliography{report} % bibliography data in report.bib

@article{Close2025,
doi = {10.3847/2041-8213/adf7a5},
url = {https://doi.org/10.3847/2041-8213/adf7a5},
year = {2025},
month = {aug},
publisher = {The American Astronomical Society},
volume = {990},
number = {1},
pages = {L9},
author = {Close, Laird M. and van Capelleveen, Richelle F. and Weible, Gabriel and Wagner, Kevin and Haffert, Sebastiaan Y. and Males, Jared R. and Ilyin, Ilya and Kenworthy, Matthew A. and Li, Jialin and Long, Joseph D. and Ertel, Steve and Ginski, Christian and Weinberger, Alycia J. and Follette, Kate and Liberman, Joshua and Twitchell, Katie and Johnson, Parker and Kueny, Jay and Apai, Daniel and Doyon, Rene and Foster, Warren and Gasho, Victor and Van Gorkom, Kyle and Guyon, Olivier and Kautz, Maggie Y. and McLeod, Avalon and McEwen, Eden and Pearce, Logan and Schatz, Lauren and Hedglen, Alexander D. and Wu, Ya-Lin and Isbell, Jacob and Power, Jenny and Carlson, Jared and Close, Emmeline and Tonucci, Elena and Mars, Matthijs},
title = {Wide {S}eparation {P}lanets in {T}ime ({WISPIT}): Discovery of a {G}ap {H}{\ensuremath{\alpha}} {P}rotoplanet {WISPIT} 2b with {Mag}{AO-X}},
journal = {The Astrophysical Journal Letters},
}

@article{Keppler2018,
	author = {{Keppler, M.} and {Benisty, M.} and {Müller, A.} and {Henning, Th.} and {van Boekel, R.} and {Cantalloube, F.} and {Ginski, C.} and {van Holstein, R. G.} and {Maire, A.-L.} and {Pohl, A.} and {Samland, M.} and {Avenhaus, H.} and {Baudino, J.-L.} and {Boccaletti, A.} and {de Boer, J.} and {Bonnefoy, M.} and {Chauvin, G.} and {Desidera, S.} and {Langlois, M.} and {Lazzoni, C.} and {Marleau, G.-D.} and {Mordasini, C.} and {Pawellek, N.} and {Stolker, T.} and {Vigan, A.} and {Zurlo, A.} and {Birnstiel, T.} and {Brandner, W.} and {Feldt, M.} and {Flock, M.} and {Girard, J.} and {Gratton, R.} and {Hagelberg, J.} and {Isella, A.} and {Janson, M.} and {Juhasz, A.} and {Kemmer, J.} and {Kral, Q.} and {Lagrange, A.-M.} and {Launhardt, R.} and {Matter, A.} and {Ménard, F.} and {Milli, J.} and {Mollière, P.} and {Olofsson, J.} and {Pérez, L.} and {Pinilla, P.} and {Pinte, C.} and {Quanz, S. P.} and {Schmidt, T.} and {Udry, S.} and {Wahhaj, Z.} and {Williams, J. P.} and {Buenzli, E.} and {Cudel, M.} and {Dominik, C.} and {Galicher, R.} and {Kasper, M.} and {Lannier, J.} and {Mesa, D.} and {Mouillet, D.} and {Peretti, S.} and {Perrot, C.} and {Salter, G.} and {Sissa, E.} and {Wildi, F.} and {Abe, L.} and {Antichi, J.} and {Augereau, J.-C.} and {Baruffolo, A.} and {Baudoz, P.} and {Bazzon, A.} and {Beuzit, J.-L.} and {Blanchard, P.} and {Brems, S. S.} and {Buey, T.} and {De Caprio, V.} and {Carbillet, M.} and {Carle, M.} and {Cascone, E.} and {Cheetham, A.} and {Claudi, R.} and {Costille, A.} and {Delboulbé, A.} and {Dohlen, K.} and {Fantinel, D.} and {Feautrier, P.} and {Fusco, T.} and {Giro, E.} and {Gluck, L.} and {Gry, C.} and {Hubin, N.} and {Hugot, E.} and {Jaquet, M.} and {Le Mignant, D.} and {Llored, M.} and {Madec, F.} and {Magnard, Y.} and {Martinez, P.} and {Maurel, D.} and {Meyer, M.} and {Möller-Nilsson, O.} and {Moulin, T.} and {Mugnier, L.} and {Origné, A.} and {Pavlov, A.} and {Perret, D.} and {Petit, C.} and {Pragt, J.} and {Puget, P.} and {Rabou, P.} and {Ramos, J.} and {Rigal, F.} and {Rochat, S.} and {Roelfsema, R.} and {Rousset, G.} and {Roux, A.} and {Salasnich, B.} and {Sauvage, J.-F.} and {Sevin, A.} and {Soenke, C.} and {Stadler, E.} and {Suarez, M.} and {Turatto, M.} and {Weber, L.}},
	title = {Discovery of a planetary-mass companion within the gap of the transition disk around {PDS} 70},
	DOI= "10.1051/0004-6361/201832957",
	url= "https://doi.org/10.1051/0004-6361/201832957",
	journal = {Astronomy \& Astrophysics},
	year = 2018,
	volume = 617,
	pages = "A44",
}

@ARTICLE{Haffert2019,
       author = {{Haffert}, S.~Y. and {Bohn}, A.~J. and {de Boer}, J. and {Snellen}, I.~A.~G. and {Brinchmann}, J. and {Girard}, J.~H. and {Keller}, C.~U. and {Bacon}, R.},
        title = "{Two accreting protoplanets around the young star PDS 70}",
      journal = {Nature Astronomy},
         year = 2019,
        month = jun,
       volume = {3},
        pages = {749-754},
          doi = {10.1038/s41550-019-0780-5},
archivePrefix = {arXiv},
       eprint = {1906.01486},
 primaryClass = {astro-ph.EP},
       adsurl = {https://ui.adsabs.harvard.edu/abs/2019NatAs...3..749H}
}

@ARTICLE{Jovanovic2023,
       author = {{Jovanovic}, Nemanja and {Gatkine}, Pradip and {Anugu}, Narsireddy and {Amezcua-Correa}, Rodrigo and {Basu Thakur}, Ritoban and {Beichman}, Charles and {Bender}, Chad F. and {Berger}, Jean-Philippe and {Bigioli}, Azzurra and {Bland-Hawthorn}, Joss and {Bourdarot}, Guillaume and {Bradford}, Charles M. and {Broeke}, Ronald and {Bryant}, Julia and {Bundy}, Kevin and {Cheriton}, Ross and {Cvetojevic}, Nick and {Diab}, Momen and {Diddams}, Scott A. and {Dinkelaker}, Aline N. and {Duis}, Jeroen and {Eikenberry}, Stephen and {Ellis}, Simon and {Endo}, Akira and {Figer}, Donald F. and {Fitzgerald}, Michael P. and {Gris-Sanchez}, Itandehui and {Gross}, Simon and {Grossard}, Ludovic and {Guyon}, Olivier and {Haffert}, Sebastiaan Y. and {Halverson}, Samuel and {Harris}, Robert J. and {He}, Jinping and {Herr}, Tobias and {Hottinger}, Philipp and {Huby}, Elsa and {Ireland}, Michael and {Jenson-Clem}, Rebecca and {Jewell}, Jeffrey and {Jocou}, Laurent and {Kraus}, Stefan and {Labadie}, Lucas and {Lacour}, Sylvestre and {Laugier}, Romain and {{\L}awniczuk}, Katarzyna and {Lin}, Jonathan and {Leifer}, Stephanie and {Leon-Saval}, Sergio and {Martin}, Guillermo and {Martinache}, Frantz and {Martinod}, Marc-Antoine and {Mazin}, Benjamin A. and {Minardi}, Stefano and {Monnier}, John D. and {Moreira}, Reinan and {Mourard}, Denis and {Nayak}, Abani Shankar and {Norris}, Barnaby and {Obrzud}, Ewelina and {Perraut}, Karine and {Reynaud}, Fran{\c{c}}ois and {Sallum}, Steph and {Schiminovich}, David and {Schwab}, Christian and {Serbayn}, Eugene and {Soliman}, Sherif and {Stoll}, Andreas and {Tang}, Liang and {Tuthill}, Peter and {Vahala}, Kerry and {Vasisht}, Gautam and {Veilleux}, Sylvain and {Walter}, Alexander B. and {Wollack}, Edward J. and {Xin}, Yinzi and {Yang}, Zongyin and {Yerolatsitis}, Stephanos and {Zhang}, Yang and {Zou}, Chang-Ling},
        title = "{2023 Astrophotonics Roadmap: pathways to realizing multi-functional integrated astrophotonic instruments}",
      journal = {Journal of Physics: Photonics},
         year = 2023,
        month = oct,
       volume = {5},
       number = {4},
          eid = {042501},
        pages = {042501},
          doi = {10.1088/2515-7647/ace869},
archivePrefix = {arXiv},
       eprint = {2311.00615},
 primaryClass = {astro-ph.IM},
       adsurl = {https://ui.adsabs.harvard.edu/abs/2023JPhP....5d2501J}
}

@article{LeonSaval2010,
author = {Sergio G. Leon-Saval and Alexander Argyros and Joss Bland-Hawthorn},
journal = {Opt. Express},
number = {8},
pages = {8430--8439},
publisher = {Optica Publishing Group},
title = {Photonic lanterns: a study of light propagation in multimode to single-mode converters},
volume = {18},
month = {Apr},
year = {2010},
url = {https://opg.optica.org/oe/abstract.cfm?URI=oe-18-8-8430},
doi = {10.1364/OE.18.008430},
}

@article{Birks2015,
author = {T. A. Birks and I. Gris-S\'{a}nchez and S. Yerolatsitis and S. G. Leon-Saval and R. R. Thomson},
journal = {Adv. Opt. Photon.},
number = {2},
pages = {107--167},
publisher = {Optica Publishing Group},
title = {The photonic lantern},
volume = {7},
month = {Jun},
year = {2015},
url = {https://opg.optica.org/aop/abstract.cfm?URI=aop-7-2-107},
doi = {10.1364/AOP.7.000107},
}

@ARTICLE{Norris2020,
       author = {{Norris}, Barnaby R.~M. and {Wei}, Jin and {Betters}, Christopher H. and {Wong}, Alison and {Leon-Saval}, Sergio G.},
        title = "{An all-photonic focal-plane wavefront sensor}",
      journal = {Nature Communications},
         year = 2020,
        month = oct,
       volume = {11},
          eid = {5335},
        pages = {5335},
          doi = {10.1038/s41467-020-19117-w},
archivePrefix = {arXiv},
       eprint = {2003.05158},
 primaryClass = {astro-ph.IM},
       adsurl = {https://ui.adsabs.harvard.edu/abs/2020NatCo..11.5335N}
}

@article{Lin2025,
author = {Jonathan Lin and Michael P. Fitzgerald and Yinzi Xin and Yoo Jung Kim and Olivier Guyon and Barnaby Norris and Christopher Betters and Sergio Leon-Saval and Kyohoon Ahn and Vincent Deo and Julien Lozi and S\'{e}bastien Vievard and Daniel Levinstein and Steph Sallum and Nemanja Jovanovic},
journal = {Opt. Lett.},
number = {8},
pages = {2780--2783},
publisher = {Optica Publishing Group},
title = {Experimental and on-sky demonstration of spectrally dispersed wavefront sensing using a photonic lantern},
volume = {50},
month = {Apr},
year = {2025},
url = {https://opg.optica.org/ol/abstract.cfm?URI=ol-50-8-2780},
doi = {10.1364/OL.551624},
}

@article{Sengupta2026,
doi = {10.3847/1538-3881/ae2617},
url = {https://doi.org/10.3847/1538-3881/ae2617},
year = {2026},
month = {jan},
publisher = {The American Astronomical Society},
volume = {171},
number = {2},
pages = {65},
author = {Sengupta, Aditya R. and Diaz, Jordan and DeMartino, Matthew and Jensen-Clem, Rebecca and Cetre, Sylvain and Gates, Elinor and Bundy, Kevin and Dillon, Daren and Hinz, Philip and Salama, Maïssa and Skaf, Nour and Guyon, Olivier and Crowe, Tara and Dobias, Caleb and Eikenberry, Stephen S. and Amezcua-Correa, Rodrigo and Yerolatsitis, Stephanos},
title = {On-sky Demonstration of Second-stage Wave-front Control with a Photonic Lantern},
journal = {The Astronomical Journal},
}

@article{Xin2022,
doi = {10.3847/1538-4357/ac9284},
url = {https://doi.org/10.3847/1538-4357/ac9284},
year = {2022},
month = {oct},
publisher = {The American Astronomical Society},
volume = {938},
number = {2},
pages = {140},
author = {Xin, Yinzi and Jovanovic, Nemanja and Ruane, Garreth and Mawet, Dimitri and Fitzgerald, Michael P. and Echeverri, Daniel and Lin, Jonathan and Leon-Saval, Sergio and Gatkine, Pradip and Kim, Yoo Jung and Norris, Barnaby and Sallum, Steph},
title = {Efficient Detection and Characterization of Exoplanets within the Diffraction Limit: Nulling with a Mode-selective Photonic Lantern},
journal = {The Astrophysical Journal},
}

@article{Xin2024,
author = {Yinzi Xin and Daniel Echeverri and Nemanja Jovanovic and Dimitri Mawet and Sergio Leon-Saval and Rodrigo Amezcua-Correa and Stephanos Yerolatsitis and Michael P. Fitzgerald and Pradip Gatkine and Yoo Jung Kim and Jonathan Lin and Barnaby Norris and Garreth Ruane and Steph Sallum},
title = {{Laboratory demonstration of a Photonic Lantern Nuller in monochromatic and broadband light}},
volume = {10},
journal = {Journal of Astronomical Telescopes, Instruments, and Systems},
number = {2},
publisher = {SPIE},
pages = {025001},
year = {2024},
doi = {10.1117/1.JATIS.10.2.025001},
URL = {https://doi.org/10.1117/1.JATIS.10.2.025001}
}

@article{Kim2025,
doi = {10.3847/2041-8213/ae0739},
url = {https://doi.org/10.3847/2041-8213/ae0739},
year = {2025},
month = {oct},
publisher = {The American Astronomical Society},
volume = {993},
number = {1},
pages = {L3},
author = {Kim, Yoo Jung and Fitzgerald, Michael P. and Vievard, Sébastien and Lin, Jonathan and Xin, Yinzi and Lucas, Miles and Guyon, Olivier and Lozi, Julien and Deo, Vincent and Huby, Elsa and Lacour, Sylvestre and Lallement, Manon and Amezcua-Correa, Rodrigo and Leon-Saval, Sergio and Norris, Barnaby and Nowak, Mathias and Sallum, Steph and Sarrazin, Jehanne and Taras, Adam and Yerolatsitis, Stephanos and Jovanovic, Nemanja},
title = {On-sky Demonstration of Subdiffraction-limited Astronomical Measurement Using a Photonic Lantern},
journal = {The Astrophysical Journal Letters},
}

@ARTICLE{Vievard2024,
       author = {{Vievard}, S. and {Lallement}, M. and {Leon-Saval}, S. and {Guyon}, O. and {Jovanovic}, N. and {Huby}, E. and {Lacour}, S. and {Lozi}, J. and {Deo}, V. and {Ahn}, K. and {Lucas}, M. and {Sallum}, S. and {Norris}, B. and {Betters}, C. and {Amezcua-Correa}, R. and {Yerolatsitis}, S. and {Fitzgerald}, M.~P. and {Lin}, J. and {Kim}, Y.~J. and {Gatkine}, P. and {Kotani}, T. and {Tamura}, M. and {Currie}, T. and {Kenchington}, H.-D. and {Martin}, G. and {Perrin}, G.},
        title = "{Spectroscopy using a visible photonic lantern at the Subaru Telescope: Laboratory characterization and the first on-sky demonstration on Ikiiki ({\ensuremath{\alpha}} Leo) and 'Aua ({\ensuremath{\alpha}} Ori)}",
      journal = {Astronomy \& Astrophysics},
         year = 2024,
        month = nov,
       volume = {691},
          eid = {A140},
        pages = {A140},
          doi = {10.1051/0004-6361/202450234},
archivePrefix = {arXiv},
       eprint = {2409.06958},
 primaryClass = {astro-ph.IM},
       adsurl = {https://ui.adsabs.harvard.edu/abs/2024A&A...691A.140V}
}

@inproceedings{Diaz2024,
author = {Jordan Diaz and Rebecca Jensen-Clem and Daren Dillon and Philip M. Hinz and Matthew C. DeMartino and Kevin Bundy and Stephen Eikenberry and Peter Delfyett and Rodrigo Amezcua-Correa},
title = {{Laboratory demonstration of an all-fiber-based focal plane nulling interferometer}},
volume = {13095},
booktitle = {Optical and Infrared Interferometry and Imaging IX},
editor = {Jens Kammerer and Stephanie Sallum and Joel Sanchez-Bermudez},
organization = {International Society for Optics and Photonics},
publisher = {SPIE},
pages = {130952N},
year = {2024},
doi = {10.1117/12.3019617},
URL = {https://doi.org/10.1117/12.3019617}
}

@ARTICLE{Betti2022,
       author = {{Betti}, S.~K. and {Follette}, K.~B. and {Ward-Duong}, K. and {Aoyama}, Y. and {Marleau}, G.-D. and {Bary}, J. and {Robinson}, C. and {Janson}, M. and {Balmer}, W. and {Chauvin}, G. and {Palma-Bifani}, P.},
        title = "{Near-infrared Accretion Signatures from the Circumbinary Planetary-mass Companion Delorme 1 (AB)b}",
      journal = {The Astrophysical Journal Letters},
         year = 2022,
        month = aug,
       volume = {935},
       number = {1},
          eid = {L18},
        pages = {L18},
          doi = {10.3847/2041-8213/ac85ef},
archivePrefix = {arXiv},
       eprint = {2208.05016},
 primaryClass = {astro-ph.EP},
       adsurl = {https://ui.adsabs.harvard.edu/abs/2022ApJ...935L..18B}
}

@ARTICLE{Marleau2024,
       author = {{Marleau}, Gabriel-Dominique and {Aoyama}, Yuhiko and {Hashimoto}, Jun and {Zhou}, Yifan},
        title = "{Revisiting the Helium and Hydrogen Accretion Indicators at TWA 27B: Weak Mass Flow at Near-freefall Velocity}",
      journal = {The Astrophysical Journal},
         year = 2024,
        month = mar,
       volume = {964},
       number = {1},
          eid = {70},
        pages = {70},
          doi = {10.3847/1538-4357/ad1ee9},
archivePrefix = {arXiv},
       eprint = {2401.04763},
 primaryClass = {astro-ph.EP},
       adsurl = {https://ui.adsabs.harvard.edu/abs/2024ApJ...964...70M}
}

@inproceedings{JensenClem2021,
author = {Rebecca Jensen-Clem and Daren Dillon and Benjamin Gerard and M.A.M. van Kooten and J. Fowler and Renate Kupke and Sylvain Cetre and Dominic Sanchez and Philip Hinz and Cesar Laguna and David Doelman and Frans Snik},
title = {{The Santa Cruz Extreme AO Lab (SEAL): design and first light}},
volume = {11823},
booktitle = {Techniques and Instrumentation for Detection of Exoplanets X},
editor = {Stuart B. Shaklan and Garreth J. Ruane},
organization = {International Society for Optics and Photonics},
publisher = {SPIE},
pages = {118231D},
year = {2021},
doi = {10.1117/12.2594676},
URL = {https://doi.org/10.1117/12.2594676}
}

@inproceedings{JensenClem2025,
author = {Rebecca Jensen-Clem and Vincent Chambouleyron and Prince Javier and Daren Dillon and Emiel H. Por and Benjamin Calvin and Sylvain Cetre and Rodrigo Amezcua Correa and Tara Crowe and Jordan Diaz and Caleb Dobias and David Doelman and Stephen Eikenberry and J. Fowler and Benjamin L. Gerard and Phil Hinz and Renate Kupke and Ashai Moreno and Tiffany Nguyen and Maissa Salama and Aditya R. Sengupta and Nour Skaf and Frans Snik},
title = {{The Santa Cruz Extreme AO Lab (SEAL) 2.0: a reflective, multiwavelength rebuild}},
volume = {13627},
booktitle = {Techniques and Instrumentation for Detection of Exoplanets XII},
editor = {Garreth J. Ruane and Maxwell A. Millar-Blanchaer},
organization = {International Society for Optics and Photonics},
publisher = {SPIE},
pages = {1362727},
year = {2025},
doi = {10.1117/12.3065350},
URL = {https://doi.org/10.1117/12.3065350}
}

@inproceedings{Lin2022a,
author = {Jonathan Lin and Sebastian Vievard and Nemanja Jovanovic and Barnaby Norris and Michael P. Fitzgerald and Christopher Betters and Pradip Gatkine and Olivier Guyon and Yoo Jung Kim and Sergio Leon-Saval and Julien Lozi and Dimitri Mawet and Steph Sallum and Yinzi Xin},
title = {{Experimental measurements of AO-fed photonic lantern coupling efficiencies}},
volume = {12188},
booktitle = {Advances in Optical and Mechanical Technologies for Telescopes and Instrumentation V},
editor = {Ram{\'o}n Navarro and Roland Geyl},
organization = {International Society for Optics and Photonics},
publisher = {SPIE},
pages = {121882E},
year = {2022},
doi = {10.1117/12.2630608},
URL = {https://doi.org/10.1117/12.2630608}
}

@INPROCEEDINGS{Tedder2019,
       author = {{Tedder}, Sarah A. and {Vyhnalek}, Brian E. and {Leon-Saval}, Sergio and {Betters}, Christopher and {Floyd}, Bert and {Staffa}, Jeremy and {Lafon}, Robert},
        title = "{Single-mode fiber and few-mode fiber photonic lanterns performance evaluated for use in a scalable real-time photon counting ground receiver}",
    booktitle = {Free-Space Laser Communications XXXI},
         year = 2019,
       editor = {{Hemmati}, Hamid and {Boroson}, Don M.},
       series = {Society of Photo-Optical Instrumentation Engineers (SPIE) Conference Series},
       volume = {10910},
        month = mar,
          eid = {109100G},
        pages = {109100G},
          doi = {10.1117/12.2507478},
       adsurl = {https://ui.adsabs.harvard.edu/abs/2019SPIE10910E..0GT}
}

@inproceedings{Sengupta2025,
author = {Aditya R. Sengupta and Vincent Chambouleyron and Jordan Diaz and Matthew DeMartino and Rebecca Jensen-Clem and Benjamin L. Gerard and Michael J. Messerly and Paul Pax and Daren Dillon and Kevin Bundy and Maria Cuevas and Sylvain Cetre and Bruce Macintosh and Caleb Dobias and Tara Crowe and Stephen S. Eikenberry and Rodrigo Amezcua-Correa and Stephanos Yerolatsitis},
title = {{Experimental validation of photonic lantern imaging and wavefront sensing performance}},
volume = {13627},
booktitle = {Techniques and Instrumentation for Detection of Exoplanets XII},
editor = {Garreth J. Ruane and Maxwell A. Millar-Blanchaer},
organization = {International Society for Optics and Photonics},
publisher = {SPIE},
pages = {136271X},
year = {2025},
doi = {10.1117/12.3064074},
URL = {https://doi.org/10.1117/12.3064074}
}

@PHDTHESIS{Hinz2021,
       author = {{Hinz}, Philip Mark},
        title = "{Nulling interferometry for studying other planetary systems: Techniques and observations}",
       school = {University of Arizona},
         year = 2001,
        month = oct,
       adsurl = {https://ui.adsabs.harvard.edu/abs/2001PhDT.........8H}
}

@book{Goodman2015,
author = {Goodman, Joseph W.},
address = {Hoboken, New Jersey},
booktitle = {Statistical optics},
edition = {Second edition.},
isbn = {9781119009481},
language = {eng},
publisher = {Wiley},
series = {Wiley Series in Pure and Applied Optics},
title = {Statistical optics },
year = {2015 - 2015},
}

@ARTICLE{Akcay2002,
       author = {{Akcay}, Ceyhun and {Parrein}, Pascale and {Rolland}, Jannick P.},
        title = "{Estimation of Longitudinal Resolution in Optical Coherence Imaging}",
      journal = {Applied Optics},
         year = 2002,
        month = sep,
       volume = {41},
       number = {25},
        pages = {5256-5262},
          doi = {10.1364/AO.41.005256},
       adsurl = {https://ui.adsabs.harvard.edu/abs/2002ApOpt..41.5256A}
}

@inproceedings{Birbacher2024,
author = {Thomas Birbacher and Adrian M. Glauser and Mohanakrishna Ranganathan and Jonah T. Hansen and Suvrath Mahadevan and Sascha P. Quanz},
title = {{Beam metrology and control for the Nulling Interferometry Cryogenic Experiment}},
volume = {13095},
booktitle = {Optical and Infrared Interferometry and Imaging IX},
editor = {Jens Kammerer and Stephanie Sallum and Joel Sanchez-Bermudez},
organization = {International Society for Optics and Photonics},
publisher = {SPIE},
pages = {1309538},
year = {2024},
doi = {10.1117/12.3018652},
URL = {https://doi.org/10.1117/12.3018652}
}

@inproceedings{Kim2024,
author = {Yoo Jung Kim and Michael P. Fitzgerald and Jonathan Lin and Julien Lozi and Sebastien Vievard and Nemanja Jovanovic and Sergio Leon-Saval and Kyohoon Ahn and Christopher Betters and Vincent Deo and Pradip Gatkine and Olivier Guyon and Manon Lallement and Daniel Levinstein and Dimitri Mawet and Barnaby Norris and Steph Sallum and Yinzi Xin},
title = {{Spectral characterization of 3-port photonic lantern for spectroastrometry}},
volume = {13095},
booktitle = {Optical and Infrared Interferometry and Imaging IX},
editor = {Jens Kammerer and Stephanie Sallum and Joel Sanchez-Bermudez},
organization = {International Society for Optics and Photonics},
publisher = {SPIE},
pages = {130950T},
year = {2024},
doi = {10.1117/12.3019017},
URL = {https://doi.org/10.1117/12.3019017}
}
\bibliographystyle{spiebib} % makes bibtex use spiebib.bst

\end{document}